\documentclass[article]{elsarticle}
\usepackage[T1]{fontenc}
\usepackage{subfig}
\usepackage{caption}
\usepackage{lineno,hyperref}
\usepackage{amsmath}
\usepackage{amssymb}
\usepackage{mathtools}
\usepackage{tablefootnote}
\usepackage{soul}
\usepackage{ulem}
\usepackage{color}
\usepackage{float}
\usepackage{makecell}
\usepackage[usenames,dvipsnames,svgnames,table]{xcolor}
\usepackage{indentfirst}
\usepackage{leftidx}
\usepackage{slashed}

\begin{document}
\begin{frontmatter}
\title{Spinors and Scalars in curved spacetime: neutrino dark energy (DE$_{\nu}$)}

\author[ICC,UB]{Ali Rida Khalifeh\corref{cor1}}
\ead{ark93@icc.ub.edu}
\cortext[cor1]{Corresponding author}

\author[ICC,ICREA]{Raul Jimenez}
\ead{raul.jimenez@icc.ub.edu}

\address[ICC]{ICC, University of Barcelona, Marti  i Franques, 1, E-08028 Barcelona, Spain.}
\address[UB]{Dept. de  Fisica Cuantica y Astrofisica, University of Barcelona, Marti  i Franques 1, E-08028 Barcelona, Spain.}
\address[ICREA]{ICREA, Pg. Lluis Companys 23, Barcelona, E-08010, Spain.}

\begin{abstract}
We study the interaction, in general curved spacetime, between a spinor and a scalar field describing dark energy; the so-called DE$_{\nu}$ model in curved space. The dominant term is the dimension 5 operator, which results in different energy shifts for the neutrino states: an Aharonov-Bohm-like effect. We study the phenomenology of this term and make observational predictions to detect dark energy interactions in the laboratory due to its effect on neutrino oscillation experiments, which opens up the possibility of designing underground experiments to detect dark energy. This dimension 5 operator beyond the Standard Model interaction is less suppressed than the widely discussed dimension 6 operator, which corresponds to mass varying neutrinos; the dimension 5 operator does not suffer from gravitational instabilities.
\end{abstract}
\end{frontmatter}

\section{Introduction}

The physical nature of the current phase of cosmic acceleration \cite{Riess:acelerating1,Perlmutter:acelerating2}, associated to an entity called Dark Energy (DE), remains a major mystery. This is  despite the fact that it has been observationally studied intensively and confirmed via  very different cosmological observables, most notably, the cosmic microwave background (CMB)~\cite{Planck2018,CMB:acelerating2},  supernovae~\cite{Riess:acelerating1,Perlmutter:acelerating2}, baryon acoustic oscillations (BAO)~\cite{BAO:acelerating}, large scale structure \cite{LSS:aceleration}, cosmic chronometers~\cite{cc1,cc2,cc3} and weak lensing~\cite{WeakLensing:acelerating}.

On the observational front, it is becoming clear that the effective equation of state of DE is compatible with a cosmological constant at the \% level, i.e. $w=-1$ where $p =w \rho$~\cite{Planck2018}, with $p$ and $\rho$ being the pressure and energy density, respectively, of DE. Given how strongly the observations suggest that DE is a cosmological constant, it is interesting to explore possible alternatives, given the large difference between the value of the energy density of a cosmological constant (which is $\sim (\text{meV})^4$) and the vacuum expectation value $M_{\rm p}^4$, where $M_{\rm p}$ is the Planck mass (see e.g. Ref.~\cite{Weinberg:LambdaProblem,Weinberg:LambdaProblem2}). Further, but not exclusive to the cosmological constant, there is the coincidence problem (see e.g., Ref.~\cite{Wang:Coincidence,Velten:Coincidence}), i.e., the fact that the redshift of equality between CDM and $\Lambda$ is close to us in time.

In order to overcome all these problems, several alternatives have been proposed. One of the first ideas was that of a dynamic DE~\cite{Ratra:rolling,Caldwell:DEevolving}, which involves a  minimally coupled dynamical scalar field (quintessence). The latter models have quickly gained popularity, as they alleviate the cosmological coincidence problem. Further extensions of quintessence  models include the addition of  a coupling with other sectors of the Universe, the so-called interacting quintessence. Most  interacting models couple DE to the other ``dark'' component of the universe, Dark Matter (DM) (see for instance Refs.~\cite{Micheletti:DMDE,Atrio:DMDE,Bohmer:EoM2, LopezH}).  However,  a coupling of the scalar field to DM in general induces effects akin to modifications of gravity beyond the simple description offered by General Relativity (GR). These modifications are being increasingly constrained by observations~\cite{1995A&A...301..321W,2008JCAP...07..020V,2009NuPhS.194..260M,2010MNRAS.402.2344M,2010MNRAS.402.2355V,2000MNRAS.312..521A,2000PhRvD..62d3511A,2004PhRvD..69l3516M,2008PhRvD..78b3015A,2012PhRvD..86j3507P,2010AIPC.1241.1016L,2010JCAP...09..029L,2013PhRvD..88b3531S,2018JCAP...06..007E,2009JCAP...07..034G}. 

This shortcoming  is avoided in a recently proposed alternative~\cite{2018PDU....20...72S},  where a generic scalar field is ``frozen'' in place by coupling with neutrinos (or any other particle, although neutrinos have several advantages; for one, we know they exist), and can thus act as DE. A coupling between neutrinos and the scalar field responsible for DE is motivated by the similarity in scale between neutrino rest-mass and the energy scale of dark energy ($\sim \text{meV}$). Another advantage of neutrinos is that they become non-relativistic at relatively recent redshifts ($z \sim 10$) thus providing a possible alieviation of the ``why now?'' problem.

The main aim of this paper is to develop the formalism in curved space, using semi-classical effective field theory, for the lowest order interaction between spinor and scalar fields in order to constrain the possible interactions that could lead to momentum (and energy) transfer as in the phenomenological model of Ref.~\cite{2018PDU....20...72S}. In order to do, so we explore all possible terms of interaction permitted by the symmetries in the standard model of particle physics (SM) to a scalar field. In this respect our model is very minimal as it only requires the current standard model, which we know exists, and one extra scalar field, the only postulated ingredient in this model. Our derivation is totally general and can be applied to any scenario in which a spinor and a scalar field interact in curved spacetime.

The structure of this paper is as follows: in \S~\ref{sec:general} we derive, within the effective field theory framework, equations of motion due to a general type of interaction between a spinor and a scalar field that leads to momentum, and energy, transfer. We explore the dynamics (using a semi-classical approach) and phenomenology (the effect on neutrino oscillations) of the 5th dimension operator in \S~\ref{sec:dim5}. We summarize our results in \S~\ref{sec:summary}. In an appendix, we solve the equations of motion of dimension operators 6 and 8 as to fully complete our analysis. Units in which $8\pi G=c=1$, and a metric signature $(-+++)$ will be used. 

\section{General Framework}
\label{sec:general}
	
	In this section, we present the general framework used to study the interaction of  a spinor field (particularly neutrino) and a scalar field in curved background. This will be applicable to any type of interaction between the two fields. In later sections we specify the type of interactions and study their consequences (see Ref.~\cite{Birrell:1982ix} for details on Spinors in curved spacetime).

    What distinguishes neutrinos from other fields in curved spacetime, is the fact that the general linear group GL(4), which is that of general coordinate transformations, does not have a spinorial representation. This inhibits the generalization of equations of motion in the standard way (substituting partial derivatives with covariant ones, and the flat metric with that of curved background), and requires the use of tetrads, as we will see shortly.
	
	The most general action for a real classical scalar field $\varphi$ and a spinor field $\psi$, with its hermitian conjugate $\psi^{\dagger}$, interacting in a curved spacetime with a metric $g_{\mu\nu}$ takes the form:
	\begin{equation}
	S=S_{\text{gravity}}+S_{\text{scalar}}+S_{\text{spinor}}+S_{\text{interaction}}
	\label{eq:FormOfAction}
	\end{equation} 
	where $S_{\text{gravity}}$ is the gravitational action, $S_{\text{scalar}}$ is that of the scalar field, $S_{\text{spinor}}$ is the one of spinor fields and $S_{\text{interaction}}$ is for the interaction term. Note that both fields are coupled minimally to gravity, as a first step in studying the dynamics in curved spacetime. More explicitly, this action takes the form: 
	\begin{equation}
		S=\int d^4x\sqrt{-g}\bigg[\frac{1}{2}R-\frac{1}{2}\mathcal{D}_{\mu}\varphi\mathcal{D}^{\mu}\varphi-V(\varphi)+i\hbar\big(\bar{\psi}\gamma^{\mu}\mathcal{D}_{\mu}\psi-\mathcal{D}_{\mu}\bar{\psi}\gamma^{\mu}\psi\big)-2m\bar{\psi}\psi+\lambda\Theta\bigg]
		\label{eq:GeneralAction}
\end{equation}
	where $g$ is the determinant of $g_{\mu\nu}$, $R=g^{\mu\nu}R_{\mu\nu}$ is the Ricci scalar, the trace of the Ricci tensor $R_{\mu\nu}$ and $\hbar$ is the reduced Planck constant. Moreover, $\mathcal{D}_{\mu}$ is the spacetime covariant derivative that takes into account the spin of the field. For instance, $\mathcal{D}_{\mu}$ reduces to a partial derivative $\partial_{\mu}$ when applied to a scalar field, or to the usual covariant derivative of GR, $\nabla_{\mu}$, when applied to a vector or tensor fields. The explicit form of the covariant derivative for a spinor field in curved spacetime will be introduced later. Furthermore, $V(\varphi)$ is the potential for the scalar field and $\bar{\psi}=\psi^{\dagger}\gamma^0$, with $\gamma^0$ being one of the Dirac gamma matrices $\gamma^{\mu}$. Finally, $m$ is the mass of the spinor field, $\lambda$ is the coupling constant between the scalar and the spinor, as described by the interaction term $\Theta(\psi,\varphi,X_{\psi},X_{\varphi})$, with $X^{\mu}_{\varphi}\big(X_{\psi}^{\mu}\big)=\nabla^{\mu}\varphi\big(\mathcal{D}^{\mu}\psi\big)$. 
	
The equation of motion for the spinor field is obtained by varying ~\eqref{eq:FormOfAction} with respect to the spinor field:
\begin{equation}
\frac{1}{\sqrt{-g}}\frac{\delta S}{\delta\bar{\psi}}=0\quad \Rightarrow\quad i\hbar\gamma^{\mu}\mathcal{D}_{\mu}\psi-m\psi=-\frac{\lambda}{2}\bigg(\frac{\partial\Theta}{\partial\bar{\psi}}-\mathcal{D}^{\mu}\frac{\partial\Theta}{\partial X^{\mu}_{\bar{\psi}}}\bigg)
\label{eq:Spinor}
\end{equation}
which will be our main focus in this work. The variation of the action with respect to $\psi$ will give the complex conjugate of~\eqref{eq:Spinor}. For completeness, we state the equation of motion for the scalar field:
\begin{equation}
\Box\varphi-\frac{\partial V}{\partial\varphi}=-\lambda\bigg(\frac{\partial\Theta}{\partial\varphi}-\nabla^{\mu}\frac{\partial\Theta}{\partial X^{\mu}_{\varphi}}\bigg).
\end{equation}
where $\Box=g^{\mu\nu}\nabla_{\mu}\nabla_{\nu}$. 

\section{Dimension $5$ operator: Linear derivative coupling}
\label{sec:dim5}

In effective field theory, the lowest order interaction term beyond the SM interactions, between spinor and scalar fields which is allowed by the SM and GR symmetries, is\footnote{Note that this interaction does not produce any gravitational instabilities, as already shown in Ref.~\cite{2018PDU....20...72S}}: 
\begin{equation}
\Theta=J^{\mu}\nabla_{\mu}\varphi
\label{eq:5thDim}
\end{equation}
 where $J^{\mu}=\bar{\psi}\gamma^{\mu}\psi$. The Dirac equation in curved spacetime~\eqref{eq:Spinor} is:
 \begin{equation}
 \big(i\hbar\gamma^{\mu}\mathcal{D}_{\mu}-m\big)\psi=-\frac{\lambda}{2}\gamma^{\mu}\psi\nabla_{\mu}\varphi.
 \label{eq:Dirac5thDim}
 \end{equation}
 If we focus on regions much smaller than the curvature scale, we can use the WKB approximation to study the dynamics of spinors in a gravitational field~\cite{Mukhanov:2007zz} (see~\ref{WKB} for a brief discussion on the WKB approximation). The strategy would be to expand the field in powers of $\hbar$, and then study the dynamics at each power. In this case, the spinor field can be written as\footnote{Note that here the phase $S(x)$ is slowly varying compared to $\psi_n$. Therefore there's no need to apply a WKB expansion on the phase.}:
 \begin{equation}
 \psi(x)=e^{iS(x)/\hbar}\sum_{n=0}^{\infty}(-i\hbar)^n\psi_n(x)
 \label{eq:WKBexpan}
 \end{equation}
where the $\psi_n$s are also spinors. Plugging this in~\eqref{eq:Dirac5thDim}, keeping terms up to first order, we get:
\begin{equation}
\bigg[-\big(\gamma^{\mu}\partial_{\mu}S+m\big)+\frac{\lambda}{2}\gamma^{\mu}\partial_{\mu}\varphi\bigg]\psi_0+i\hbar\bigg[\bigg(\gamma^{\mu}\partial_{\mu}S+m-\frac{\lambda}{2}\gamma^{\mu}\partial_{\mu}\varphi\bigg)\psi_1+\gamma^{\mu}\mathcal{D}_{\mu}\psi_0\bigg]=0.
\label{eq:5DimDiracWKB}
\end{equation}

\subsection{Solution at 0${}^{\text{th}}$ order in WKB expansion}

From~\eqref{eq:5DimDiracWKB}, we can read off the 0${}^{\text{th}}$ order equation to be:
\begin{equation}
\big(\gamma^{\mu}\partial_{\mu}S+m\big)\psi_0=\frac{\lambda}{2}\gamma^{\mu}\partial_{\mu}\varphi\psi_0.
\label{eq:5Dim0thOrder}
\end{equation}
A non-trivial solution for this algebraic set of equations exists if
\begin{equation}
\text{det}\bigg[\gamma^{\mu}\partial_{\mu}\bigg(S-\frac{\lambda}{2}\varphi\bigg)+m\bigg]=0\quad\Rightarrow\quad\partial_{\mu}\bigg(S-\frac{\lambda}{2}\varphi\bigg)\partial^{\mu}\bigg(S-\frac{\lambda}{2}\varphi\bigg)=-m^2,
\end{equation}
which is the Hamilton-Jacobi equation for a spinless relativistic particle. Therefore, its canonical 4-momentum and 4-velocities are defined as:
\begin{equation}
p^{\alpha}=\partial^{\alpha}\bigg(S-\frac{\lambda}{2}\varphi\bigg);\quad u^{\alpha}=\frac{p^{\alpha}}{m}
\label{eq:5DimMomentum}
\end{equation}
giving the usual normalizations:
\begin{equation}
p^{\alpha}p_{\alpha}=-m^2; \quad u^{\alpha}u_{\alpha}=-1.
\label{eq:Normalization}
\end{equation}
Notice that if we calculate the vorticity $\omega_{\alpha\beta}=\frac{1}{2}\big(\nabla_{\alpha}u_{\beta}-\nabla_{\beta}u_{\alpha}\big)$ by direct substitution of~\eqref{eq:5DimMomentum} we find that it's 0. Hence, at 0${}^{\text{th}}$ order, the spinor field is equivalent to an irrotational fluid of spin 0 particles. This means that these particles follow the geodesic equation without alteration:
\begin{equation}
u^{\alpha}\nabla_{\alpha}u^{\beta}=0\quad\Rightarrow\quad \frac{dp^{\alpha}}{d\tau}+\frac{1}{m}\Gamma^{\alpha}_{\ \beta\gamma}p^{\beta}p^{\gamma}=0
\end{equation}
where $\tau$ is the proper time of the particle. This result is consistent with the findings of Ref.~\cite{2018PDU....20...72S}, and it will be at every order in $\hbar$, as one can check by simply noticing that the equation of the scalar field does not change. Indeed the latter is:
\begin{equation}
g^{\mu\nu}\nabla_{\mu}\nabla_{\nu}\varphi-\frac{\partial V}{\partial\varphi}=-\lambda\bigg(\frac{\partial\Theta}{\partial\varphi}-\nabla^{\mu}\frac{\partial\Theta}{\partial X^{\mu}_{\varphi}}\bigg)=-\lambda\mathcal{D}_{\mu}J^{\mu}=0
\end{equation}
where the last equality follows from~\eqref{eq:Dirac5thDim} and its complex conjugate.

Although this type of interactions doesn't affect the dynamics, it still causes a shift in the energy of the neutrinos, as has been claimed previously in Ref.~\cite{2018PDU....20...72S}. To see this quantitatively, consider the Lagrangian density for neutrinos:
\begin{equation}
\mathcal{L}_{\nu}=i\hbar\big(\bar{\psi}\gamma^{\mu}\mathcal{D}_{\mu}\psi-\mathcal{D}_{\mu}\bar{\psi}\gamma^{\mu}\psi\big)-2m\bar{\psi}\psi+\lambda\bar{\psi}\gamma^{\mu}\psi\partial_{\mu}\varphi.
\end{equation}
The conjugate momentum of the field would be:
\begin{equation}
\pi_{\nu}=\pi_{\psi}+\pi_{\bar{\psi}}=\frac{\delta\mathcal{L}_{\nu}}{\delta\mathcal{D}_t\psi}+\frac{\delta\mathcal{L}_{\nu}}{\delta\mathcal{D}_t\bar{\psi}}
\end{equation}
and therefore the Hamiltonian density would be:
\begin{equation}
\mathcal{H}=\pi_{\psi}\mathcal{D}_t\psi+\pi_{\bar{\psi}}\mathcal{D}_t\bar{\psi}-\mathcal{L}=i\hbar\big(\bar{\psi}\vec{\gamma}.\vec{\mathcal{D}}\psi-\vec{\mathcal{D}}\bar{\psi}.\vec{\gamma}\psi\big)+2m\bar{\psi}\psi-\lambda\bar{\psi}\gamma^{\mu}\psi\partial_{\mu}\varphi.
\end{equation}
The last term is an additional contribution to the neutrino energy that comes from this interaction. If we consider a homogeneous and isotropic scalar field, i.e $\varphi=\varphi(t)$, then at 0${}^{\text{th}}$ order in WKB expansion, this term would be of the form $\bar{n}_{\nu}\dot{\varphi}$, where $\bar{n}_{\nu}$ is the average number density of the neutrino particles, and a dot denotes derivative with respect to cosmic time. More interestingly, if the neutrino fluid is moving with a bulk velocity in a gravitational potential well, the additional term would take the form $\lambda\bar{n}_{\nu}\big(\dot{\varphi}+\vec{v}.\vec{\nabla}\varphi\big)$, where $\vec{v}$ is the bulk velocity. Of course this effect will be at perturbation level if we are considering a homogeneous and isotropic scalar field.

Let us see now in more detail the effect of this shift in energy on neutrino oscillations, and constrain the coupling $\lambda$ to get an observable effect.

\subsection{Phenomenology of 5${}^{\text{th}}$ dim Operator: Effect on Neutrino Oscillations}
\label{NeutrinoOscillSection}
When studying neutrino oscillations, it is customary to write the neutrino state in terms of mass eigenstates and spacetime coordinates, as done for instance in Refs.~\cite{PhysRevD.24.110,Giunti:2007ry}. However, since we are considering curved spacetime, it would be better to write things in a covariant way~\cite{PhysRevD.55.7960}:
\begin{equation}
|\Psi_{\alpha}(\lambda)\rangle=\sum_{j}^{}U_{\alpha j}e^{i\int_{\lambda_0}^{\lambda}\vec{P}.\vec{q}d\lambda'}|\nu_j\rangle
\label{eq:NeutrinoState}
\end{equation}
where $|\Psi_{\alpha}\rangle$ is the neutrino state that was initially in a flavor $\alpha$ and  $\lambda$ is the affine parameter that characterizes the neutrino's world-line, with $\lambda_0$ its value today. Moreover, $U_{\alpha j}$ is the conversion matrix between flavor and mass eigenstates, $\vec{P}$ is the 4-momentum operator (generating spacetime translations) of the mass eigenstates $|\nu_j\rangle$ and $\vec{q}=d\vec{x}/d\lambda$ is a null vector tangent to the neutrino's world-line $\vec{x}(\lambda)=\big[t(\lambda), x(\lambda), y(\lambda), z(\lambda)\big]$. If we concentrate on transitions between electron neutrinos, $\nu_e$, and muon neutrinos, $\nu_{\mu}$, we can define a vector of transition amplitudes:
\begin{align}
\chi(\lambda) &= \begin{bmatrix}
\langle\nu_e|\Psi(\lambda)\rangle \\
\langle\nu_{\mu}|\Psi(\lambda)\rangle
\end{bmatrix}
\end{align}
which satisfies the differential equation:
\begin{equation}
i\frac{d\chi}{d\lambda}=\vec{P}.\vec{q}\ \chi,
\label{eq:AmplitudeEvolution}
\end{equation} 
 with the solution given in~\eqref{eq:NeutrinoState}. Our goal is therefore to calculate the quantity $\vec{P}.\vec{q}$ for neutrinos traveling in curved spacetime and interacting with the scalar field $\varphi$, with an interaction given by~\eqref{eq:5thDim}. To this end, let us rewrite the Dirac equation~\eqref{eq:Dirac5thDim} for a column vector of neutrino flavors $\psi_f$ (we consider two neutrino flavors for simplicity):
\begin{equation}
\bigg[i\hbar\bigg(\gamma^{\mu}\mathcal{D}_{\mu}-\frac{i}{\hbar}A_{\varphi\mu}\mathcal{P}_L\bigg)-M_f\bigg]\psi_f=0
\end{equation} 
where
\begin{align}
A_{\varphi\mu} &=-\frac{1}{2}\partial^{\mu}\varphi \begin{pmatrix}
\lambda_e & 0 \\
0 & \lambda_{\mu}
\end{pmatrix}
\end{align}
and we are considering different coupling constants for the two flavors $\nu_e$ and $\nu_{\mu}$. Moreover, $M_f$ is the vacuum mass matrix in flavor space, given by:
\begin{align}
M_f^2 &= U\begin{pmatrix}
m_1^2 & 0 \\
0 & m_2^2
\end{pmatrix}U^{\dagger}
\label{eq:MassMatrix}
\end{align}
where
\begin{align}
U &= \begin{pmatrix}
\cos\theta & \sin\theta \\
-\sin\theta & \cos\theta
\end{pmatrix}
\end{align}
is the mixing matrix, with mixing angle $\theta$, that transforms from one basis to another, and $m_1$ and $m_2$ are eigenvalues for mass eigenstates. Finally, $\mathcal{P}_L=\frac{1}{2}(1-\gamma^5)$ is the left-handed projection operator. From now on we will focus on left-handed neutrinos only, and therefore drop this factor. Furthermore, the explicit form of the covariant derivative is~\cite{10.1088/978-1-627-05330-3}
\begin{equation}
\gamma^{\mu}\mathcal{D}_{\mu}=\gamma^ae_a^{\mu}\big(\partial_{\mu}+\Gamma_{\mu}\big)
\end{equation}
where $\gamma^a$ are the Dirac matrices in local inertial coordinates, $e_a^{\mu}$ are tetrad (or vierbein) fields that connect general coordinates to local ones, and
\begin{equation}
\Gamma_{\mu}=\frac{1}{8}\big[\gamma^b,\gamma^c\big]e_b^{\nu}\nabla_{\mu}e_{c\nu}
\end{equation} 
is the spin connection that describes the effect of gravity on the spin of the particle, with $\big[\gamma^a,\gamma^b\big]$  being the commutator of the two matrices $\gamma^a$ and $\gamma^b$. We adopt the convention that Latin indices correspond to local inertial coordinates, while Greek ones correspond to general coordinates. From here, it can be shown that
\begin{equation}
\gamma^ae_a^{\mu}\Gamma_{\mu}=\frac{i}{\hbar}\gamma^ae_a^{\mu}A_{G\mu}
\end{equation} 
where
\begin{equation}
A_{G}^{\mu}=\frac{1}{4}\sqrt{-g}e_a^{\mu}\epsilon^{abcd}(\partial_{\sigma}e_{b\nu}-\partial_{\nu}e_{b\sigma})e_c^{\nu}e_d^{\sigma}
\end{equation}
with $\epsilon^{abcd}$ being the Levi-civita symbol in four dimensions. Hence, the Dirac equation will take the form:
\begin{equation}
\bigg[i\hbar\gamma^{\mu}\bigg(\partial_{\mu}-\frac{i}{\hbar}A_{\mu}\bigg)-M_f\bigg]\psi_f=0
\end{equation}
with $A^{\mu}=A_G^{\mu}+A_{\varphi}^{\mu}$. For this equation to have a non-trivial solution, the mass-shell relation must be satisfied, i.e:
\begin{equation}
\big(P^{\mu}+A^{\mu}\big)\big(P_{\mu}+A_{\mu}\big)=M_f^2.
\label{eq:MassShell}
\end{equation}
We already know what $\vec{q}$ is and that it satisfies the geodesic equation. Since we want to find $\vec{P}.\vec{q}$, we can construct $\vec{P}$ as done in~\cite{PhysRevD.55.7960}: First, take neutrinos to be energy eigenstates, that is $P^0=q^0$. Second, assume $\vec{P}$ and $\vec{q}$ to be parallel (we don't really need the perpendicular component, since we are taking inner product of the two vectors in the end), which means we can write $P^i=q^i(1-\epsilon)$. Finally, for relativistic neutrinos, $\epsilon<<1$, and therefore~\eqref{eq:MassShell} gives:
\begin{equation}
\vec{P}.\vec{q}=-\epsilon\big(g_{0i}q^0q^i+g_{ij}q^iq^j\big)=\frac{1}{2}M_f^2-q^{\mu}A_{\mu}.
\end{equation} 
We can now write~\eqref{eq:NeutrinoState} as
\begin{equation}
|\Psi_{\alpha}(\lambda)\rangle=\sum_{j}U_{\alpha j}e^{i\Omega}|\nu_j\rangle
\end{equation}
where
\begin{equation}
\Omega=\int_{\lambda_0}^{\lambda}\bigg(\frac{1}{2}M_f^2-q^{\mu}A_{\mu}\bigg)d\lambda'.
\end{equation}
For a flat FRW universe:
\begin{equation}
ds^2=-dt^2+a^2\delta_{ij}dx^idx^j.
\label{eq:FRW}
\end{equation}
A convenient choice for the tetrad fields is:
\begin{equation}
e_a^{\mu}=\text{diag}\big[1,a^{-1},a^{-1},a^{-1}\big]
\end{equation}
from which one can show, after some algebra, that $A_G^{\mu}$=0. Physically, this can be seen as a consequence of having a homogeneous and isotropic spacetime, and therefore there would be no alteration in the spin direction. Furthermore, using~\eqref{eq:FRW}, one can show that if neutrinos are traveling along null trajectories, which is convenient when studying oscillations, then the affine parameter is related to cosmic time by:
\begin{equation}
dt=Ed\lambda
\end{equation}
where $E=q^0$ is the neutrino energy. From here, we can write~\eqref{eq:AmplitudeEvolution} as
\begin{equation}
i\frac{d\chi}{dt}=-\frac{1}{2E}\bigg(M_f^2+V_{\varphi}\bigg)\chi
\end{equation}
where
\begin{align}
V_{\varphi}=-2q^{\mu}A_{\varphi\mu} &=E\dot{\varphi}\begin{pmatrix}
\lambda_e & 0\\
0 & \lambda_{\mu}
\end{pmatrix} 
\end{align}
and we have take into account that, at the background level, $\varphi=\varphi(t)$. On the other hand, when written explicitly from~\eqref{eq:MassMatrix}, 
\begin{align}
M_f^2 &=\begin{pmatrix}
m_1^2+\Delta\sin^2\theta & \frac{1}{2}\Delta\sin 2\theta\\
\frac{1}{2}\Delta\sin 2\theta & m_1^2 +\Delta\cos^22\theta
\end{pmatrix}\nonumber \\
&= \bigg(m_1^2+\frac{1}{2}\Delta\bigg)I+\frac{1}{2}\Delta\begin{pmatrix}
-\cos 2\theta & \sin 2\theta\\
\sin 2\theta & \cos 2\theta
\end{pmatrix}
\end{align}
where $\Delta=m_2^2-m_1^2$ and $I$ is the identity matrix. The term proportional to the identity matrix in the above will be a common phase factor for both transition amplitudes, hence we can ignore it for oscillation purposes. Moreover, if we want to measure this effect on Earth, we can consider distances small enough for us to safely assume a Minkowsky spacetime, in which case we can write $dt\approx dx$. In addition, to detect the effect of this interaction on neutrino oscillations, we look at the difference in frequency $\Omega$ between the presence of this interaction and its absence:
\begin{equation}
\Delta\Omega=\Omega-\Omega_{\text{NoInt}}=\frac{1}{2E}\int_{x0}^{x}V_{\varphi}dx'
\end{equation}
where $\Omega_{\text{NoInt}}$ is the frequency without interactions and $x_0$ is the position of Earth. So, for a specific flavor $i$,
\begin{equation}
\Delta\Omega_i=\frac{\lambda_i}{2E}\int_{x0}^{x}E\dot{\varphi}\ dx'.
\end{equation}
Let's assume, for order of magnitude estimate purposes, that $E$ and $\dot{\varphi}$ are roughly constants. Therefore
\begin{equation}
\lambda_i\sim \frac{\Delta\Omega_i}{\dot{\varphi}\Delta x}
\end{equation}
where $\Delta x$ is the distance traveled by the neutrino between its interaction and detection points. This will give us the order of magnitude of the interaction parameter depending on the nature of the scalar field under consideration. For instance, if $\varphi$ is supposed to describe DE, then its energy scale is $\mathcal{O}(\text{meV})$, hence $\dot{\varphi}\sim 10^{-8}\text{eV}^2$. If the distance traveled is of the size of our galaxy (around 50kpc), to get a difference in frequency $\mathcal{O}(\pi)$, then $\lambda_i\sim 10^{-9}$GeV${}^{-1}$. On the other hand, if the scalar field is the Higgs, which means its energy scale is $\mathcal{O}(100\text{GeV})$, then $\dot{\varphi}\sim 10^{20}\text{eV}^2$, and so $\lambda_i\sim 10^{-37}\text{GeV}^{-1}$. It is more likely therefore that this interaction to be relevant for DE rather than the Higgs. In other words, the DE cannot be the Higgs field in this model, but we need to postulate an extra scalar field. We will discuss elsewhere signatures of this models in specific neutrino oscillation experiments like IceCube. Here we only note that the effect is in principle measurable.

In the above discussion, we haven't seen a direct effect on the equations of motion due to the curved gravitational field. This will be apparent at 1${}^{\text{st}}$ order in the WKB expansion.

\subsection{Solution at 1${}^{\text{st}}$ order in WKB expansion}

From~\eqref{eq:5DimDiracWKB}, we can read off the 1${}^{\text{st}}$ order equation of motion to be
\begin{equation}
\bigg(\gamma^{\mu}\partial_{\mu}\bigg(S-\frac{\lambda}{2}\varphi\bigg)+m\bigg)\psi_1=-\gamma^{\mu}\mathcal{D}_{\mu}\psi_0.
\label{eq:5Dim1stOrder}
\end{equation}
Since this is a non-homogeneous linear algebraic equation, the solutions of the homogeneous equation for a Hermitian system, which is $\psi_0$, should be orthogonal to the inhomogeneity, i.e\footnote{This equation can be proved directly from the complex conjugate of~\eqref{eq:5Dim0thOrder} and~\eqref{eq:5Dim1stOrder}.}
\begin{equation}
\bar{\psi}_0\gamma^{\mu}\mathcal{D}_{\mu}\psi_0=0,
\label{eq:Orthogonal}
\end{equation}
 in order to insure non-trivial solutions at 1${}^{\text{st}}$ order. This relation can be used to show that
 \begin{equation}
 u^{\alpha}\mathcal{D}_{\alpha}\psi_0=-\frac{\theta}{2}\psi_0
 \end{equation}
 where $\theta=\mathcal{D}_{\alpha}u^{\alpha}$, which is equivalent to saying that $\psi_0$ follows a sourced geodesic in curved spacetime. Moreover, for later convenience, define a spinor $\xi_0$ such that 
 \begin{equation}
\psi_0=f(x)\xi_0,
\label{eq:DefXi}
 \end{equation} 
 where $f(x)$ is a function of the coordinates. Therefore, the above relation translates into:
 \begin{equation}
 u^{\alpha}\partial_{\alpha}f=-\frac{\theta}{2}f; \quad u^{\alpha}\mathcal{D}_{\alpha}\xi_0=0.
 \end{equation} 
 As we will see shortly, these relations are useful when calculating the deviation from the 0${}^{\text{th}}$ order geodesic motion due to curvature of spacetime. To this end, let's start by noticing that the Dirac current, $J^{\mu}=\bar{\psi}\gamma^{\mu}\psi$ can be decomposed into convection and magnetization currents\footnote{This relation can be derived by starting from the definition of the magnetization current and using the Dirac equation}:
 \begin{equation}
 J^{\mu}=J^{\mu}_{c}+J^{\mu}_M
 \end{equation}
 where
 \begin{equation}
 J^{\mu}_{c}=-\frac{\hbar}{2mi}\big[\big(\tilde{\mathcal{D}}^{\mu}\bar{\psi}\big)\psi-\bar{\psi}\tilde{\mathcal{D}}^{\mu}\psi\big]; \quad J^{\mu}_M=\frac{\hbar}{2m}\tilde{\mathcal{D}}_{\nu}\big(\bar{\psi}\sigma^{\mu\nu}\psi\big)
 \end{equation}
 are the convective and magnetization currents, respectively, with $\tilde{\mathcal{D}}_{\nu}=\mathcal{D}_{\nu}-i\frac{\lambda}{2}\partial_{\nu}\varphi$ and $\sigma^{\mu\nu}=\frac{i}{2}[\gamma^{\mu},\gamma^{\nu}]$. Using the WKB expansions~\eqref{eq:WKBexpan} and~\eqref{eq:DefXi}, we get the convection current to be, to 1${}^{\text{st}}$ order in $\hbar$:
 \begin{equation}
 J^{\mu}_c=f^2\bigg[u^{\mu}-\frac{\hbar}{2mi}\bigg(\tilde{\mathcal{D}}^{\mu}\bar{\xi}_0\xi_0-\bar{\xi}_0\tilde{\mathcal{D}}^{\mu}\xi_0\bigg)\bigg]+\mathcal{O}(\hbar^2).
 \end{equation}
 Moreover, the convection current describes the probability flow of a particle moving with a velocity $v^{\mu}$, i.e it's proportional to the latter, which is
 \begin{equation}
 v^{\mu}=u^{\mu}-\frac{\hbar}{2mi}\big(\tilde{\mathcal{D}}^{\mu}\bar{\xi}_0\xi_0-\bar{\xi}_0\tilde{\mathcal{D}}^{\mu}\xi_0\big).
 \end{equation}
 Therefore the deviation from geodesic motion at order $\hbar$ is
 \begin{equation}
 \delta u^{\mu}=\frac{\hbar}{2mi}\big(\tilde{\mathcal{D}}^{\mu}\bar{\xi}_0\xi_0-\bar{\xi}_0\tilde{\mathcal{D}}^{\mu}\xi_0\big)=\frac{\hbar}{2mi}\big(\mathcal{D}^{\mu}\bar{\xi}_0\xi_0-\bar{\xi}_0\mathcal{D}^{\mu}\xi_0\big)
 \end{equation}
 where the last equality shows that the linear derivative coupling has no effect on the dynamics, as expected. Note that this deviation from geodesic motion can be interpreted in terms of an additional force due to the spin-curvature coupling, which can be written as
 \begin{equation}
 f^{\mu}=m\frac{\mathcal{D}v^{\mu}}{\mathcal{D}\tau}=mv^{\nu}\mathcal{D}_{\nu}v^{\mu}=\frac{\hbar}{4}g^{\mu\nu}u^{\alpha}R_{\nu\alpha\gamma\delta}\bar{\xi}_0\sigma^{\gamma\delta}\xi_0
 \end{equation}
 where $R_{\nu\alpha\gamma\delta}$ is the Riemann curvature tensor (see~\cite{10.1088/978-1-627-05330-3} for details about the derivation). What this means is that there will be a force due to the interaction of the spinor neutrino field with gravity, at order $\hbar$. This force will result in a change of the neutrino momentum that appears in section~\ref{NeutrinoOscillSection}, and therefore will affect the resulting neutrino oscillations. Furthermore, this additional force will change the scaling of the momentum with the scale factor. However, observationally, it would be difficult to detect such a change with current technologies due to the fact that this extra term is $\mathcal{O}$($\hbar$) smaller than the other terms in the geodesic equation. We will leave the details of this result for future work. Note also that this force will not alter the dynamics of the scalar field for two reasons: first, this force exists irrespective of whether there is an interaction between the spinor and the scalar fields (as we've shown above), and second, as the scalar field is a classical field, such an effect would not alter it's motion.
 
 This concludes the results for the linear derivative coupling between the neutrino spinor and a scalar field. As we can see, this type of interaction affects the energy density of the spinor field, but does not alter the dynamics. The latter change due to the spin-curvature coupling at order $\hbar$.

\section{Discussion and Summary}
\label{sec:summary}

We have studied the interactions between spinor and scalar fields in curved spacetime, respecting all symmetries allowed by the SM of particle physics. We have studied the most  dominant interaction beyond the SM ones in a semi-classical manner, using the WKB approximation. This term is the 5${}^{\text{th}}$ dimension interaction which causes a shift in the energy of neutrinos. This shift is similar to the Aharonov-Bohm effect, as the one described qualitatively in section 5 of the DE$_{\nu}$ model~\cite{2018PDU....20...72S}, and therefore we were able to confirm this quantitatively. We have studied the phenomenology of this effect on neutrinos oscillations and provided a test for underground laboratories to detect this interaction. This could open the possibility of detecting dark energy in the laboratory.

\section*{Acknowledgments}
Funding for this work was partially provided by the Spanish ministry of science under project PGC2018-098866-B- I00. We thank Carlos Pena-Garay for useful discussions and feedback. We also thank the anonymous referee for a very useful referee report.
 
 \appendix
 \section{WKB Approximation}
 \label{WKB}
 
 The Wentzel, Kramers, Brillouin(WKB) approximation is a method for obtaining a global approximation to the solution of a linear differential equation whose highest derivative is multiplied by a small parameter~\cite{Bender}. In Quantum Mechanics, it is usually used to solve for the wave function of the Shr$\ddot{\text{o}}$dinger equation in regions where the wavelength is much smaller than the typical distance over which the potential energy varies~\cite{Sakurai:2011zz}. This is the key requirement for the applicability of the WKB approximation, which allows one then to assume a solution $\psi(x)$ of the form
 \begin{equation}
 \psi(x)=e^{if(x)/\hbar}
 \label{eq:SolWKB}
 \end{equation}
 where $f(x)$ is a complex function. By expanding $f(x)$ in powers of $\hbar$, plugging in the Shr$\ddot{\text{o}}$dinger equation, and solving at each level in powers of $\hbar$, one can then get an approximate solution to the problem considered(see, for instance, problem 8.2 in~\cite{Griffiths}. The 0$^{\text{th}}$ order solution will give the classical solution to the Hamilton-Jacobi equation, which shows why the WKB method is a semi-classical approximation. For the situation discussed in our paper, the same concept applies, where we focus on regions such that the wavelength is much smaller than the typical distance over which the curvature varies(in the case of an FLRW universe, we would be interested in cases where $k>>H$, with $k$ being the wavenumber). As a generalization of~\ref{eq:SolWKB}, one can use the solution presented earlier in this work, eq.~\ref{eq:WKBexpan}, as has been done in~\cite{Reity:2001dr,Reity:2003ze,VanOrden:2005zw,Bolte Keppeler(1999)}.
 
 \section{6${}^{\text{th}}$dimension operator: Non-Linear Coupling}
 \label{sec:dim6}
 
 In these two appendixes we give some details on the sub-dominant interaction terms to the 5th-dimension one. The purpose is to provide details for other sub-dominant physical effects that can occur at different epochs.
 
 Consider the case where the coupling is 
 \begin{equation}
 \Theta=i\hbar\bar{\psi}\gamma^{\mu}\mathcal{D}_{\mu}\psi\varphi^2.
 \end{equation}
 Inserting this in~\eqref{eq:Spinor}, we get
 \begin{equation}
 \big(i\hbar\slashed{\mathcal{D}}-m\big)\psi=\frac{i\hbar\lambda}{2}\slashed{\mathcal{D}}\psi\varphi^2
 \end{equation}
 where $\slashed{\mathcal{D}}=\gamma^{\mu}\mathcal{D}_{\mu}$, a notation that applies to any slashed 4-vector. Note that this coupling will be an order of magnitude weaker than the dimension 5 for the same coupling constant, and so its effect on neutrino oscillations would be suppressed. Applying the WKB approximation~\eqref{eq:WKBexpan} to this equation gives, up to order $\hbar$:
 \begin{equation}
 \bigg[\bigg(1-\frac{\lambda\varphi^2}{2}\bigg)\slashed{\partial}S+m\bigg]\psi_0-i\hbar\bigg\{\bigg[\bigg(1-\frac{\lambda\varphi^2}{2}\bigg)\slashed{\partial}S+m\bigg]\psi_1+\bigg(1-\frac{\lambda}{2}\varphi^2\bigg)\slashed{\mathcal{D}}\psi_0\bigg\}=0
 \label{eq:6DimDiracWKB}
 \end{equation}
 from which we can start our analysis at each order in $\hbar$.
 \subsection{Solution at order $\hbar^0$}
 
 We can read off the equation of motion at this order from~\eqref{eq:6DimDiracWKB} to be
 \begin{equation}
 \bigg[\bigg(1-\frac{\lambda\varphi^2}{2}\bigg)\slashed{\partial}S+m\bigg]\psi_0=0.
 \label{eq:6Dim0thOrder}
 \end{equation}
  The evolution equation of the 4-momentum along the world line would be
 \begin{equation}
 \frac{dp^{\alpha}}{d\tau}+\frac{1}{m}\Gamma^{\alpha}_{\ \beta\gamma}p^{\beta}p^{\gamma}=-\frac{\lambda\varphi}{m\bigg(1-\frac{\lambda\varphi^2}{2}\bigg)}\big(m^2X^{\alpha}_{\varphi}+p^{\alpha}p_{\beta}X^{\beta}_{\varphi}\big).
 \label{eq:6DimGeodesic}
 \end{equation}
 Out of curiosity, if we multiply~\eqref{eq:6DimGeodesic} by $m$ and define $m_{\text{eff}}^2=2m^2\ln\big(1-\lambda\varphi^2/2\big)$,~\eqref{eq:6DimGeodesic} becomes
 \begin{equation}
 m\frac{dp^{\alpha}}{d\tau}+\Gamma^{\alpha}_{\ \beta\gamma}p^{\beta}p^{\gamma}=m_{\text{eff}}\frac{dm_{\text{eff}}}{d\varphi}\tilde{\partial}^{\alpha}\varphi
 \end{equation} 
 where $\tilde{\partial}^{\alpha}=\partial^{\alpha}+u^{\alpha}u_{\beta}\partial^{\beta}$, which can be interpreted as resulting from a boost in spacetime. This result is very similar to the one coming from mass varying neutrinos~\cite{Fardon_2004, PhysRevD.73.083515}, thus we will not delve much into it in detail. Before going to the order $\hbar$ solution, let us study the consequences of this interaction in a flat FRW universe
 \subsubsection{Solution at order $\hbar^0$ in a flat FRW universe}
 
 Consider the metric of spacetime~\eqref{eq:FRW}.
  The 0${}^{\text{th}}$ component of~\eqref{eq:6DimGeodesic} gives
 \begin{equation}
 \frac{1}{p}\frac{dp}{dt}+\frac{1}{a}\frac{da}{dt}=-\frac{1}{1-\frac{\lambda\varphi^2}{2}}\frac{d}{dt}\bigg(1-\frac{\lambda\varphi^2}{2}\bigg),
 \label{eq:6DimGeoCosmicTime}
 \end{equation}
 where we have used the fact that $p^0=E=\sqrt{p^2+m^2}$ and therefore $dp^0/d\tau=(p/m)dp/dt$. Note again, due to homogeneity and isotropy, at the background level, $\varphi=\varphi(t)$. The solution for~\eqref{eq:6DimGeoCosmicTime} is simply
 \begin{equation}
 p=\frac{p_0}{\tilde{a}};\quad \tilde{a}=a\bigg(1-\frac{\lambda\varphi^2}{2}\bigg).
 \end{equation}
 where $p_0$ is a positive integration constant. This result shows a shift in the momentum of the neutrino when approximated as a classical particle with spin 0. This shift involves a $\varphi^2$ term, which is similar to a mass term for the scalar field. Moreover, since $p$ is a non-negative quantity, this means that $\lambda\varphi^2<2$ in our units. This also avoids a divergence in the amplitude of the momentum.
 
 We can use this result to see the effect this interaction has on neutrino decoupling and matter-radiation equality redshift. If we assume that our effective field theory approach can be extended to energies $\mathcal{O}(1\text{MeV})$, then, since at those energies neutrinos are still relativistic, and that for a relativistic particle $p\propto T$, where $T$ is the temperature, then:
 \begin{equation}
 \frac{T_{\nu}}{T_{\gamma}}=\bigg(\frac{8}{11}\bigg)^{1/3}\bigg(1-\frac{\lambda\varphi^2}{2}\bigg)^{-1},
 \end{equation} 
 with $T_{\nu}$ and $T_{\gamma}$ are the temperatures of neutrinos and photons, respectively. Note that the factor $(8/11)^{1/3}$ appears instead of the usual $(4/11)^{1/3}$ is because, at this level in our WKB expansion, neutrinos are approximated as spin 0 particles, therefore Bosons. This result will then change the radiation content today, to be:
 \begin{equation}
 \Omega_{r0}=\Omega_{\gamma0}\bigg(1+N_{\nu}\bigg(\frac{8}{11}\bigg)^{1/3}\bigg(1-\frac{\lambda\varphi^2}{2}\bigg)^{-1}\bigg),
 \end{equation}
where $\Omega_{r0}$ and $\Omega_{\gamma0}$ are the radiation and photon density parameters today, respectively, which are explicitly defined as $\Omega_i=8\pi G\rho_i/3H_0^2$ for a specie $i$ with energy density $\rho_i$. Matter-radiation equality occurs when
\begin{equation}
\frac{\Omega_{r0}}{a_{eq}^4}=\frac{\Omega_{m0}}{a_{eq}^3},
\end{equation}
with $\Omega_{m0}$ being the density parameter of matter today, and $a_{eq}$ is the scale factor at equilibrium. This gives the redshift at matter-radiation equality to be:
\begin{equation}
1+z_{eq}=\frac{\Omega_{m0}}{\Omega_{\gamma0}}\bigg[1+N_{\nu}\bigg(\frac{8}{11}\bigg)^{1/3}\bigg(1-\frac{\lambda\varphi^2}{2}\bigg)^{-1}\bigg]^{-1}.
\end{equation}
If we use the latest Planck results~\cite{Planck2018} for the density parameters, $z_{eq}$ and $N_{\nu}$, we find $\lambda\varphi^2/2\sim\mathcal{O}(1)$.
 
 \subsection{Solution at order $\hbar^1$}
 
 At this order, from~\eqref{eq:6DimDiracWKB}, the equation of motion is:
 \begin{equation}
 \bigg[\bigg(1-\frac{\lambda\varphi^2}{2}\bigg)\slashed{\partial}S+m\bigg]\psi_1=-\bigg(1-\frac{\lambda\varphi^2}{2}\bigg)\slashed{\mathcal{D}}\psi_0
 \label{eq:6dim1stOrder}
 \end{equation}
 which can be used along with the complex conjugate of~\eqref{eq:6Dim0thOrder} to find that, also with this type of coupling, $\psi_0$ satisfies~\eqref{eq:Orthogonal}. However, when written as in~\eqref{eq:DefXi}, the equation that $f(x)$ satisfies is slightly altered:
 \begin{equation}
 u^{\alpha}\partial_{\alpha}f=-\frac{1}{2}\bigg(1-\frac{\lambda\varphi^2}{2}\bigg)\tilde{\theta}f;\quad \tilde{\theta}=\nabla_{\alpha}\bigg[\bigg(1-\frac{\lambda\varphi^2}{2}\bigg)^{-1}u^{\alpha}\bigg]
 \end{equation}
 while the one for $\xi_0(x)$ is still the same.
 
 As has been done for the case of the 5${}^{\text{th}}$ dimensional operator, we divide the Dirac current into convection and magnetization ones, and find the former to be in this case:
 \begin{equation}
 J^{\mu}=f^2\bigg[u^{\mu}+\frac{\hbar}{2mi}\bigg(\bar{\xi}_0\tilde{\mathcal{D}}^{\mu}\xi_0-\tilde{\mathcal{D}}^{\mu}\bar{\xi}_0\xi_0\bigg)\bigg]+\mathcal{O}(\hbar^2)
 \end{equation}
 but now $\tilde{\mathcal{D}}^{\mu}=\bigg(1-\frac{\lambda\varphi^2}{2}\bigg)\mathcal{D}^{\mu}$. From here, the velocity would be:
 \begin{equation}
 v^{\mu}=u^{\mu}+\frac{\hbar}{2mi}\bigg(\bar{\xi}_0\tilde{\mathcal{D}}^{\mu}\xi_0-\tilde{\mathcal{D}}^{\mu}\bar{\xi}_0\xi_0\bigg)
 \end{equation}
 and therefore the force that will alter the motion of the  $\hbar^0$ order would be:
 \begin{align}
 \frac{f^{\mu}}{m}=&\frac{\hbar}{4m}u^{\nu}g^{\mu\alpha}\bigg(1-\frac{\lambda\varphi^2}{2}\bigg)R_{\alpha\nu\gamma\delta}\bar{\xi}_0\sigma^{\gamma\delta}\xi_0+\frac{u^{\nu}g^{\mu\alpha}}{\bigg(1-\frac{\lambda\varphi^2}{2}\bigg)}\lambda\varphi\big(\partial_{\alpha}\varphi\delta u_{\nu}-\partial_{\nu}\varphi\delta u_{\alpha}\big) \nonumber
 \\
 &
 +\frac{\delta u^{\beta}g^{\mu\alpha}}{2\bigg(1-\frac{\lambda\varphi^2}{2}\bigg)}\lambda\varphi\big(u_{\alpha}\partial_{\beta}\varphi-\partial_{\alpha}\varphi u_{\beta}\big)
 \end{align}
 where 
 \begin{equation}
 \delta u^{\mu}= \frac{\hbar}{2mi}\bigg(\bar{\xi}_0\tilde{\mathcal{D}}^{\mu}\xi_0-\tilde{\mathcal{D}}^{\mu}\bar{\xi}_0\xi_0\bigg).
 \end{equation}
 
 We can see the difference between this interaction and that of the 5${}^{\text{th}}$ dimension operator. Because the former does alter the dynamics of the species involved, this alteration is manifested as well at first order in WKB, albeit in a slightly complicated way. Note also that there will be no divergence in this force, as we can see from the definition of $\tilde{\mathcal{D}}^{\mu}$. 
 
 We will now consider the last possible operator beyond the SM. We will focus only on the order $\hbar^0$ solution and its implications on the dynamics.
 
 \section{8 dimensional operator: Non-Linear derivative coupling}
 \label{sec:dim8}
 
 Consider the case where
 \begin{equation}
 \Theta=i\hbar\bar{\psi}\slashed{\mathcal{D}}\psi\partial_{\mu}\varphi\partial^{\mu}\varphi\end{equation}
 which gives the equation of motion:
 \begin{equation}
 \big(i\hbar\slashed{\mathcal{D}}-m\big)\psi=\frac{i\hbar\lambda}{2}\psi\partial_{\mu}\varphi\partial^{\mu}\varphi.
 \end{equation}
 One can see that the result is very similar to the one in the previous section, under the substitution of $\varphi^2$ with $\partial_{\mu}\varphi\partial^{\mu}\varphi$. Therefore we will avoid repeating the procedure explained above, and restrict to listing the final relevant results. 
 
 At $\hbar^0$ order, the spinor follows:
 \begin{equation}
 \bigg[\bigg(1-\lambda/2\partial_{\mu}\varphi\partial^{\mu}\varphi\bigg)\slashed{\partial}S+m\bigg]\psi_0=0
 \end{equation}
 and therefore the resulting 4-momentum will take the form:
 \begin{equation}
 p^{\alpha}=\big(1-\lambda/2\partial_{\mu}\varphi\partial^{\mu}\varphi\big)\partial^{\alpha}S.
 \end{equation}
 Again, with this type of interactions, the vorticity would be non-zero:
 \begin{equation}
 \omega_{\alpha\beta}=\frac{\lambda X_{\varphi}^{\gamma}}{2m\big(1-\lambda/2(X_{\varphi})_{\delta}X_{\varphi}^{\delta}\big)}\big[\nabla_{\beta}X_{\varphi}^{\gamma}p_{\alpha}-\nabla_{\alpha}X_{\varphi}^{\gamma}p_{\beta}\big].
 \end{equation}
 This results in the following evolution equation for the 4-momentum:
 \begin{equation}
 \frac{dp^{\alpha}}{d\tau}+\frac{1}{m}\Gamma^{\alpha}_{\ \beta\gamma}p^{\beta}p^{\gamma}=-\frac{\lambda (X_{\varphi})_{\gamma}}{m\big(1-\lambda/2(X_{\varphi})_{\delta}X_{\varphi}^{\delta}\big)}\big(m^2g^{\alpha\beta}+p^{\alpha}p^{\beta}\big)\nabla_{\beta}X_{\varphi}^{\gamma}.
 \end{equation}
 As we did for the 6${}^{\text{th}}$ dimension operator, we can multiply the above by $m$ and define an $m_{\text{eff}}^2=2m^2\ln\big[1-\lambda X_{\varphi}^{\gamma}(X_{\varphi})_{\gamma}/2\big]$ to get
 \begin{equation}
 m\frac{dp^{\alpha}}{d\tau}+\Gamma^{\alpha}_{\ \beta\gamma}p^{\beta}p^{\gamma}=m_{\text{eff}}\frac{dm_{\text{eff}}}{dX_{\varphi}^{\gamma}}\tilde{\nabla}^{\alpha}X_{\varphi}^{\gamma}
 \end{equation}
 where $\tilde{\nabla}^{\alpha}=\nabla^{\alpha}+u^{\alpha}u^{\beta}\nabla_{\beta}$ (boost like operator). This again can be interpreted in terms of mass varying neutrinos, however this time the variation is coming from a kinetic term, i.e thermal motion, while in the 6${}^{\text{th}}$ dimension case it was due to a potential term of the scalar field. If we now study the dynamics in a flat FRW universe~\eqref{eq:FRW}, we find the evolution equation for the amplitude of the momentum in cosmic time to be
 \begin{equation}
 \frac{1}{p}\frac{dp}{dt}+\frac{1}{a}\frac{da}{dt}=-\frac{1}{1+\frac{\lambda\dot{\varphi}^2}{2}}\frac{d}{dt}\bigg(1+\frac{\lambda\dot{\varphi}^2}{2}\bigg),
 \end{equation}
 and the solution 
 \begin{equation}
 p=\frac{p_0}{\tilde{a}};\quad \tilde{a}=a\big(1+\lambda\dot{\varphi}^2/2\big).
 \end{equation}
 The same shift in the evolution of the momentum is happening here as in the 6 dimensional operator, but now the shift is due to a kinetic-like term. The existence of this kinetic term allows us to interpret this redshift in the momentum of the neutrino as being due to the thermal motion of the scalar "particles".



\typeout{get arXiv to do 4 passes: Label(s) may have changed. Rerun}

\end{document}